\pdfoutput=1
\documentclass[a4paper, 12pt]{article}
\usepackage[left=2cm,right=2cm,top=2cm,bottom=2cm]{geometry}
\usepackage[english]{babel}
\usepackage{chemformula} 
\usepackage[T1]{fontenc} 
\usepackage{comment}
\usepackage{siunitx}
\usepackage{amsmath, amssymb}
\usepackage{booktabs}
\DeclareSIUnit\bar{bar}
\DeclareSIUnit\angstrom{\text {Å}}
\usepackage{url}
\usepackage[backend=biber,sorting=none,url=false,isbn=false,doi=false,style=phys]{biblatex}
\usepackage[affil-it]{authblk}
\usepackage{float}

\author[1]{Jonathan Spring}
\affil[1]{Physik-Institut, University of Zurich, 8057 Zurich, Switzerland}

\author[2]{Natalya Fedorova}
\affil[2]{Materials Research and Technology Department, Luxembourg Institute of Science and Technology, L-4362, Esch/Alzette, Luxembourg}

\author[3]{Alexandru B. Georgescu}
\affil[3]{Department of Chemistry, Indiana University, Bloomington, Indiana 47405, United States}

\author[4,5]{Alexander Vogel}
\affil[4]{Electron Microscopy Center, Empa–Swiss Federal Laboratories for Materials Science and Technology, 8600 Dübendorf, Switzerland}
\affil[5]{Swiss Nanoscience Institute, University of Basel, 4056 Basel, Switzerland}

\author[6]{Gabriele De Luca}
\affil[6]{Institut de Ciència de Materials de Barcelona (ICMAB-CSIC), 08193 Bellaterra (Barcelona), Spain}

\author[1]{Simon Jöhr}

\author[7]{Cinthia Piamonteze}
\affil[7]{Swiss Light Source, Paul Scherrer Institut, 5232 Villigen, Switzerland}

\author[8]{Marta D. Rossell}
\affil[8]{Electron Microscopy Center, Empa–Swiss Federal Laboratories for Materials Science and Technology, 8600 Dübendorf, Switzerland}

\author[2,9]{Jorge {\'{I}}{\~{n}}iguez-Gonz{\'{a}}lez}
\affil[9]{Department of Physics and Materials Science, University of Luxembourg, L-4422 Belvaux, Luxembourg}

\author[10]{Marta Gibert}
\affil[10]{Institute of Solid State Physics, TU Wien, 1040 Vienna, Austria}

\title{Engineering the Magnetic Transition Temperatures and the Rare Earth Exchange Interaction in Oxide Heterostructures}

\begin{document}
\maketitle

\newpage
\begin{abstract}
The properties of functional oxide heterostructures are strongly influenced by the physics governing their interfaces. Modern deposition techniques allow us to accurately engineer the interface physics through the growth of atomically precise heterostructures. This enables minute control over the electronic, magnetic, and structural characteristics. Here, we investigate the magnetic properties of tailor-made superlattices employing the ferromagnetic and insulating double perovskites \ch{RE2NiMnO6} (RE = La, Nd), featuring distinct Curie temperatures. Adjusting the superlattice periodicity at the unit cell level allows us to engineer their magnetic phase diagram. Large periodicity superlattices conserve the individual para- to ferromagnetic transitions of the \ch{La2NiMnO6} and \ch{Nd2NiMnO6} parent compounds. As the superlattice periodicity is reduced, the Curie temperatures of the superlattice constituents converge and, finally, collapse into one single transition for the lowest period samples. This is a consequence of the magnetic order parameter propagating across the superlattice interfaces, as supported by a minimal Landau theory model. Further, we find that the Nd-Ni/Mn exchange interaction can be enhanced by the superlattice interfaces. This leads to a field-induced reversal of the Nd magnetic moments, as confirmed by synchrotron X-ray magnetic circular dichroism measurements and supported by first-principles calculations. Our work demonstrates how superlattice engineering can be employed to fine-tune the magnetic properties in oxide heterostructures and broadens our understanding of magnetic interfacial effects.
\end{abstract}

\newpage
\section{Introduction}
Heterostructure engineering offers a powerful pathway to drastically alter the physical properties of materials compared to their bulk form or design novel functionalities. Generating artificially layered materials requires high-quality, atomically defined building blocks and precise synthesis techniques. Understanding the principles governing interface- and dimensionality-driven properties is essential for the rational design of artificially layered materials. These concepts apply to a wide range of materials, including semiconductors,\cite{Alferov1998} 2D van der Waals materials,\cite{Novoselov2016, Geim2013} and oxide materials.\cite{Hwang2012, Zubko2011, Huang2018} This work is concerned with the latter category. In their simplest form, oxide heterostructures consist of a film grown epitaxially on a single crystal substrate. The film's properties can, for example, be manipulated through strain engineering, \cite{Catalan2000, Tiwari2002, Liu2010, Catalano2014} interfacial charge transfer,\cite{Ohtomo2004, Reyren2007} dimensionality effects,\cite{Kumah2014, Golalikhani2018} or, in the case of perovskites, coupling of the oxygen octahedra rotations.\cite{Liao2016, Kan2016} The complexity of an oxide heterostructure can be further increased by raising the number of interfaces, i.e., the engineering of artificial superlattices. In addition to the specific physics governing an interface, the superlattice periodicity and the thickness of the individual layers can have significant effects on the macroscopic properties.\cite{Ramesh2019, Bhattacharya2014} This can, for example, be exploited to engineer multiferroicity, i.e., the simultaneous presence of ferroelectricity and magnetic order,\cite{Mundy2016} or to study the length scales of electronic and magnetic transitions using superlattice periodicity as a design parameter.\cite{Dominguez2020, Dominguez2023} Superlattices can also be designed such that interfacial charge transfer induces ferromagnetism, which is not present in the constituent materials.\cite{Takahashi2001, Grutter2016, Flint2017} All of the examples mentioned rely on precise growth control. Mastering the superlattice deposition with atomic precision is a prerequisite to harvesting the full potential of engineering their physical properties and functionalities.

This article investigates the magnetic properties of oxide superlattices featuring the ferromagnetic and insulating double perovskites \ch{RE2NiMnO6} (RE = rare earth). These compounds are characterized by a rock-salt ordering of the B-site Ni/Mn sublattice. Charge transfer between the ordered Ni and Mn ions leads to a valence state of \ch{Ni^{2+}}/\ch{Mn^{4+}}. According to the Goodenough-Kanamori rules, the resulting $d^8$-$d^3$ electronic configuration is associated with ferromagnetic superexchange.\cite{Blasse1965, kanamori1959} For our study, we choose the compounds \ch{La2NiMnO6} (LNMO) and \ch{Nd2NiMnO6} (NNMO) with respective bulk Curie temperatures ($T_\text{C}$) of \SI{\sim280}{\kelvin} and \SI{\sim200}{\kelvin}\cite{Dass2003, Booth2009}. Previously, we showed that bulk-like properties can be achieved in thin films and that ferromagnetism is sustained as the film thickness is reduced to a few unit cells.\cite{DeLuca2021, DeLuca2022, Spring2023} Here, we combine the LNMO and NNMO compounds into precisely engineered superlattices with atomic growth control. Having complete control over the superlattice periodicity allows us to accurately manipulate their magnetic properties. The effect is twofold. First, we use sample periodicity to fine-tune the para- to ferromagnetic phase transitions of the superlattice constituents. For the lowest periodicities ($\lesssim$\SI{2}{\nano\meter}), they combine into a single transition due to propagation of the magnetic order parameter across the interfaces, which is consistent with a simple Landau theory model. Second, the superlattice interfaces can be used to enhance the $4f$-$3d$ exchange interaction between the Nd and the Ni/Mn sublattices. This causes a field-induced reversal of the rare earth magnetic moments not present in the parent compounds. This is investigated through an extensive synchrotron study and first-principles calculations.

\section{Results and Discussion}
\subsection{Superlattice Growth and Structural Characterization}
\begin{figure*}
\centering
\includegraphics[width=\textwidth]{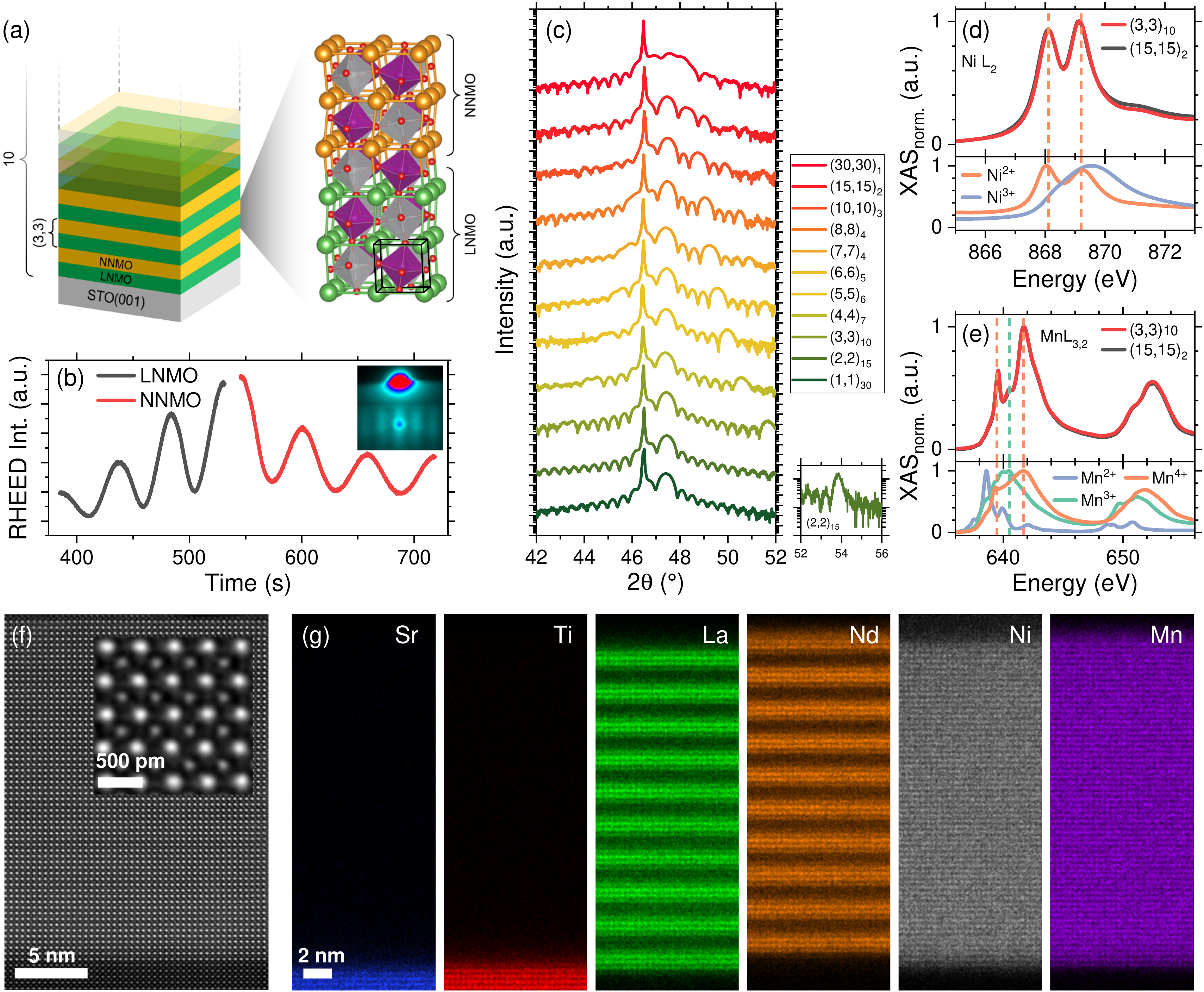}
\caption{Structural and electronic characterization. (a) Schematic of a $(3,3)_{10}$ LNMO/NNMO superlattice grown on STO(001). The pseudocubic unit cell is overlaid in black. (b) RHEED intensity vs.\ time during the growth of a $(3,3)_{10}$ superlattice. Interrupting the growth at the maximum of a RHEED oscillation allows us to switch between the two superlattice components, LNMO and NNMO, upon completion of one monolayer. The inset shows the diffraction pattern at the end of the growth. (c) XRD around the STO (002) substrate peak for superlattices of different periodicities featuring finite-size Laue oscillations and superlattice satellite peaks. (d,e) XAS around the Ni $L_{2}$ and Mn $L_{3,2}$ absorption edges for a $(3,3)_{10}$ and a $(15,15)_{2}$ superlattice in the top panels. The bottom panels show reference spectra for different Ni and Mn oxidation states. (f,g) Cross-sectional STEM imaging of a $(3,3)_{10}$ superlattice imaged along the (100)\textsubscript{pc} direction. The HAADF-STEM data in (f) highlights the structural quality of our samples. The dark region at the bottom corresponds to the STO(001) substrate, and the inset shows a magnified area of the superlattice. The EDX elemental maps in (g) reveal the distinct (3,3) superlattice layering.}
\label{fig:1}
\end{figure*}

The LNMO/NNMO superlattices are grown by radio frequency (RF) off-axis magnetron sputtering with in situ reflection high-energy electron diffraction (RHEED). The sample periodicity is varied while the total thickness is fixed to 60 pseudocubic (pc) unit cells (uc) (\SI{\sim 23}{\nano\meter}). The superlattices are denoted as $(x,x)_y$, where $x$ is the number of LNMO or NNMO pc unit cells in each layer, and the subscript $y$ indicates the number of repetitions of the LNMO/NNMO double layer. This is illustrated for a $(3,3)_{10}$ superlattice in Figure \ref{fig:1}(a). To keep the total thickness constant, the relation $2x\times y = \SI{60}{uc}$ has to be fulfilled. For $x$ values where this is impossible, the closest possible total thickness to 60 uc is chosen. In this work, $x$ is varied from 1 to 30 uc. All superlattices are grown on single crystal \ch{SrTiO3}(001) (STO) substrates with \ch{TiO2}-termination. Using the pc lattice parameters of \SI{3.881}{\angstrom} and \SI{3.845}{\angstrom} for LNMO and NNMO, respectively, the STO substrate imposes tensile strains of \SI{0.6}{\percent} and \SI{1.5}{\percent} in the respective layers. Our custom-built, RHEED-enabled sputtering setup allows us to control the superlattice periodicity in situ by switching the deposition between the LNMO and NNMO sputter sources at exactly the maximum of a RHEED intensity oscillation, Figure \ref{fig:1}(b). Further, the RHEED diffraction pattern in the inset, recorded after the completed superlattice growth, is indicative of a 2D surface. This is also confirmed by atomic force microscopy (AFM) topography. AFM images for superlattices of different periodicity are presented in SI Figure S1.

In Figure \ref{fig:1}(c), we present X-ray diffraction (XRD) scans around the STO (002) substrate peak for the entire series of superlattices of different periodicities. The measurement for the $(3,3)_{10}$ sample is shown in more detail in SI Figure S2(a). Intense Laue oscillations attest to a very high crystal quality independent of superlattice periodicity. Further, superlattice satellite peaks emerge due to sharp interfaces between the LNMO and NNMO layers. They are visible down to ultra-low periodicities of only 2 uc; see inset in Figure \ref{fig:1}(c). The B-site rock salt ordering of the Ni/Mn sublattice is probed by XRD along (111)\textsubscript{pc}. In this direction, the alternating \ch{Ni^{2+}} and \ch{Mn^{4+}} planes lead to a doubling of the periodicity and the emergence of a diffraction condition at (\nicefrac{1}{2} \nicefrac{1}{2} \nicefrac{1}{2})\textsubscript{pc}, as is shown in SI Figure S2(b) for a $(3,3)_{10}$ superlattice. Coherent strain of the superlattices was confirmed by reciprocal space map (RSM) around the STO (103) substrate peak, SI Figure S2(c). Concerning the orientation of the superlattices, RSMs reveal the emergence of a half-integer peak at (1 0 \nicefrac{5}{2})\textsubscript{pc}. This indicates a doubling of the unit cell in the sample's out-of-plane direction, which is associated with an in-phase tilt pattern in this direction. Hence, the long axis of the monoclinic $P2_1/n$ unit cell is established in the out-of-plane direction, corresponding to an $a^-a^-c^+$ tilt pattern, SI Figure S2(d-f). In the in-plane direction, we find a coexistence of domains with the monoclinic $a$ and $b$ axis along both the STO (110) and (1-10) directions, SI Figure S2(g,h). The two domains emerge because the cubic STO(001) substrate does not impose a preferred in-plane orientation.

To assess the electronic structure of the Ni/Mn sublattice, we present X-ray absorption spectroscopy (XAS) at the Ni $L_{2}$ and Mn $L_{3,2}$ edges recorded on two representative $(3,3)_{10}$ and $(15,15)_{2}$ samples, Figure \ref{fig:1}(d,e) top panels. Comparison to the respective reference spectra in the bottom panels confirms a valence state of \ch{Ni^{2+}}/\ch{Mn^{4+}} independent of periodicity. This is indicative of a double perovskite with a well-ordered Ni/Mn sublattice \cite{Blasse1965, DeLuca2021, Spring2023}.

We further characterize our samples via cross-sectional scanning transmission electron microscopy (STEM) of an example $(3,3)_{10}$ superlattice along the (100)\textsubscript{pc} direction in Figures \ref{fig:1}(f,g). The high-angle annular dark-field (HAADF) STEM data in panel (f) highlights the crystal quality and the absence of structural defects. With this imaging technique, commonly referred to as $Z$-contrast imaging because its contrast is roughly proportional to the square of the atomic number $Z$, we cannot distinguish between the LNMO and NNMO layers because the La ($Z = 57$) and Nd ($Z = 60$) atomic numbers are too close to each other. The energy dispersive X-ray (EDX) elemental maps in panel (g), on the other hand, clearly show the pronounced (3,3) La/Nd layering of the perovskite A-sublattice and the continuous Ni/Mn B-sublattice. Note that the Ni/Mn rock salt ordering is not accessible along the imaged (100)\textsubscript{pc} direction. It is, however, well established through the half-order XRD peak at (\nicefrac{1}{2} \nicefrac{1}{2} \nicefrac{1}{2})\textsubscript{pc} and the \ch{Ni^{2+}}/\ch{Mn^{4+}} valence state.\cite{DeLuca2021}

\subsection{Engineering the Magnetic Phase Diagram}
\begin{figure*}
\centering
\includegraphics[width=\textwidth]{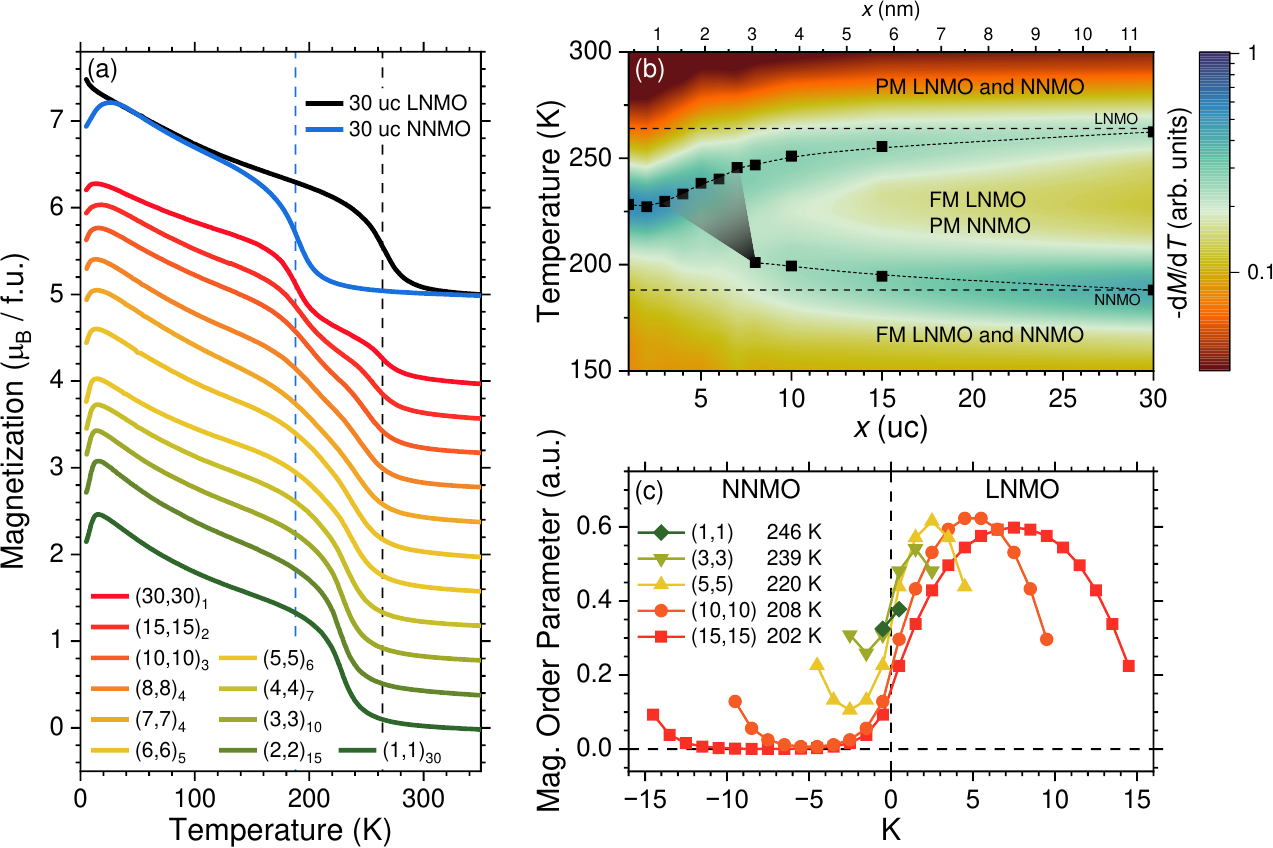}
\caption{Temperature dependence of the superlattice magnetization. (a) $M(T)$ for LNMO/NNMO superlattices of different periodicities measured at an in-plane applied field of \SI{0.5}{\tesla} while cooling from \SI{350}{\kelvin} to \SI{5}{\kelvin}. For comparison, the $M(T)$ curves for 30 uc pure LNMO and NNMO films are added. The vertical lines indicate the local minima in slope ($\nicefrac{\text{d}M}{\text{d}T}$) for the pure thin films, which are used as transition temperatures $T_\text{C}$. Note that this definition leads to $T_\text{C}$ values \SI{\sim 20}{\kelvin} lower than the standard values. The curves are offset for better visibility. (b) Phase diagram showing $\nicefrac{\text{d}M}{\text{d}T}$ as a function of superlattice period $x$ and temperature. The transition temperatures indicated by black squares are extracted from Pseudo-Voigt fits of the local minima in $\nicefrac{\text{d}M}{\text{d}T}$. Horizontal dashed lines indicate the $T_\text{C}$s of the 30 uc pure LNMO and NNMO films as shown in (a). The color scale is interpolated from the individual $\nicefrac{\text{d}M}{\text{d}T}$ curves. (c) Simulated magnetization resolved by layer in superlattices of different periodicities. For each sample, the temperature is chosen to be \SI{10}{\kelvin} lower than the high-temperature transition as determined by the minimum in the simulated $\nicefrac{\text{d}M}{\text{d}T}$ data.}
\label{fig:2}
\end{figure*}

We investigate the influence of superlattice periodicity on the para- to ferromagnetic phase transitions through SQUID magnetometry. Figure \ref{fig:2}(a) shows the magnetization per formula unit (f.u.) of (La$_2$/Nd$_2$)NiMnO$_6$ vs.\ temperature for all superlattice periodicities measured at an in-plane applied magnetic field of \SI{0.5}{\tesla} while cooling from \SI{350}{\kelvin} to \SI{5}{\kelvin}. Additionally, we present data for two pure 30 uc LNMO and NNMO thin films. Using the high-temperature onset in $M(T)$ as the Curie temperature ($T_\text{C}$), they both display bulk-like values of \SI{\sim280}{\kelvin} and \SI{\sim200}{\kelvin}, respectively \cite{Dass2003, Booth2009}. In this work, however, we use the local minimum in $\nicefrac{\text{d}M}{\text{d}T}$ to define $T_\text{C}$. This allows for a consistent determination of $T_\text{C}$ in the superlattices. Compared to the more conventional approach described above, this leads to $T_\text{C}$ values reduced by \SI{\sim 20}{\kelvin}. The respective temperatures for the LNMO and NNMO films are marked by vertical dashed lines. Returning to the superlattices, the highest periodicity $(30,30)_{1}$ sample shows two distinct magnetic transitions that correspond well to the transitions in the two pure thin films. As the superlattice periodicity is reduced, the transitions converge and eventually combine in the lowest periodicity samples. We also note a low-temperature downturn in magnetization for the superlattices and the NNMO thin film, which we will return to in the next section.

We now use the $M(T)$ data to construct a temperature vs.\ periodicity phase diagram in Figure \ref{fig:2}(b). The color scale is interpolated from the individual $\nicefrac{\text{d}M}{\text{d}T}$ curves. The black squares in the phase diagram mark the respective $T_\text{C}$s, which are determined by fitting Pseudo-Voigt functions to the $\nicefrac{\text{d}M}{\text{d}T}$ data. SI Figure S3 shows the measured data together with the fitted curves. Two distinct peaks can be fitted for superlattices with periodicity (8,8) and larger, corresponding to the LNMO and NNMO Curie temperatures. For smaller periodicities, only one $\nicefrac{\text{d}M}{\text{d}T}$ peak is found, suggesting a unique transition. However, a second transition is still visible as a shoulder in $\nicefrac{\text{d}M}{\text{d}T}$, preventing clearly determining the periodicity where they collapse into a single transition. This is represented by the gradient triangle in Figure \ref{fig:2}(b).

To shed more light on the interaction of the magnetic transitions of the LNMO and NNMO layers, we now employ Landau theory calculations. We use a second-order, one-dimensional Landau model with periodic boundary conditions, see Methods. Half of the material is treated as LNMO and the other half as NNMO. The respective transition temperatures are extracted from the $M(T)$ measurements of the pure LNMO and NNMO thin films. Figure \ref{fig:2}(c) shows the calculated magnetic order parameter resolved by layer for superlattices of different periodicities. For each sample, the temperature is chosen \SI{10}{\kelvin} below the calculated high-temperature transition in the respective structure in order to facilitate the comparison. The corresponding $\nicefrac{\text{d}M}{\text{d}T}$ data is presented in SI Figure S4, and the chosen temperatures are listed in the legend of Figure \ref{fig:2}(c). In the case of completely decoupled magnetic systems, at the chosen temperatures, LNMO would be completely ferromagnetic, while NNMO would remain paramagnetic. However, Figure \ref{fig:2}(c) reveals a pronounced propagation of the magnetic order parameter across the LNMO/NNMO interface. In the high-periodicity (15,15) superlattice, the bulk of the LNMO layer displays a strong magnetic ordering, which is drastically reduced towards the interfaces to NNMO. The ordering propagates partially into the NNMO layer (2-3 uc) before it reaches zero inside the NNMO bulk. This suggests that even in a (15,15) superlattice, the transitions are not completely separated, as can also be seen in SI Figure S4. The $\nicefrac{\text{d}M}{\text{d}T}$ curve for the NNMO layer shows a small peak at the high-temperature LNMO transition. On the other end of the spectrum, looking at the (1,1) and (3,3) samples, we still find unequal magnetization in LNMO and NNMO. Although the SQUID-measured $M(T)$ data in Figure \ref{fig:2}(a) and the calculated $\nicefrac{\text{d}M}{\text{d}T}$ in SI Figure S4 indicate one single transition, our Landau modeling strongly suggest that even at the lowest periodicities, the magnetic order parameter is not constant across the superlattice layers.

\subsection{Tuning the Exchange Interaction}
The magnetic behavior of the superlattices is mainly dominated by the ferromagnetic \ch{Ni^{2+}}/\ch{Mn^{4+}} sublattice. In contrast to the LNMO layers, the NNMO layers also contain magnetic \ch{Nd^{3+}} ions on the A-site. In pure NNMO thin films, we have shown that Nd constitutes a paramagnetic system that does not interact with its ferromagnetic Ni/Mn environment \cite{Spring2023}. How does the magnetic response of the Nd sublattice change if non-magnetic La layers are interspersed?

\begin{figure*}
\centering
\includegraphics[width=\textwidth]{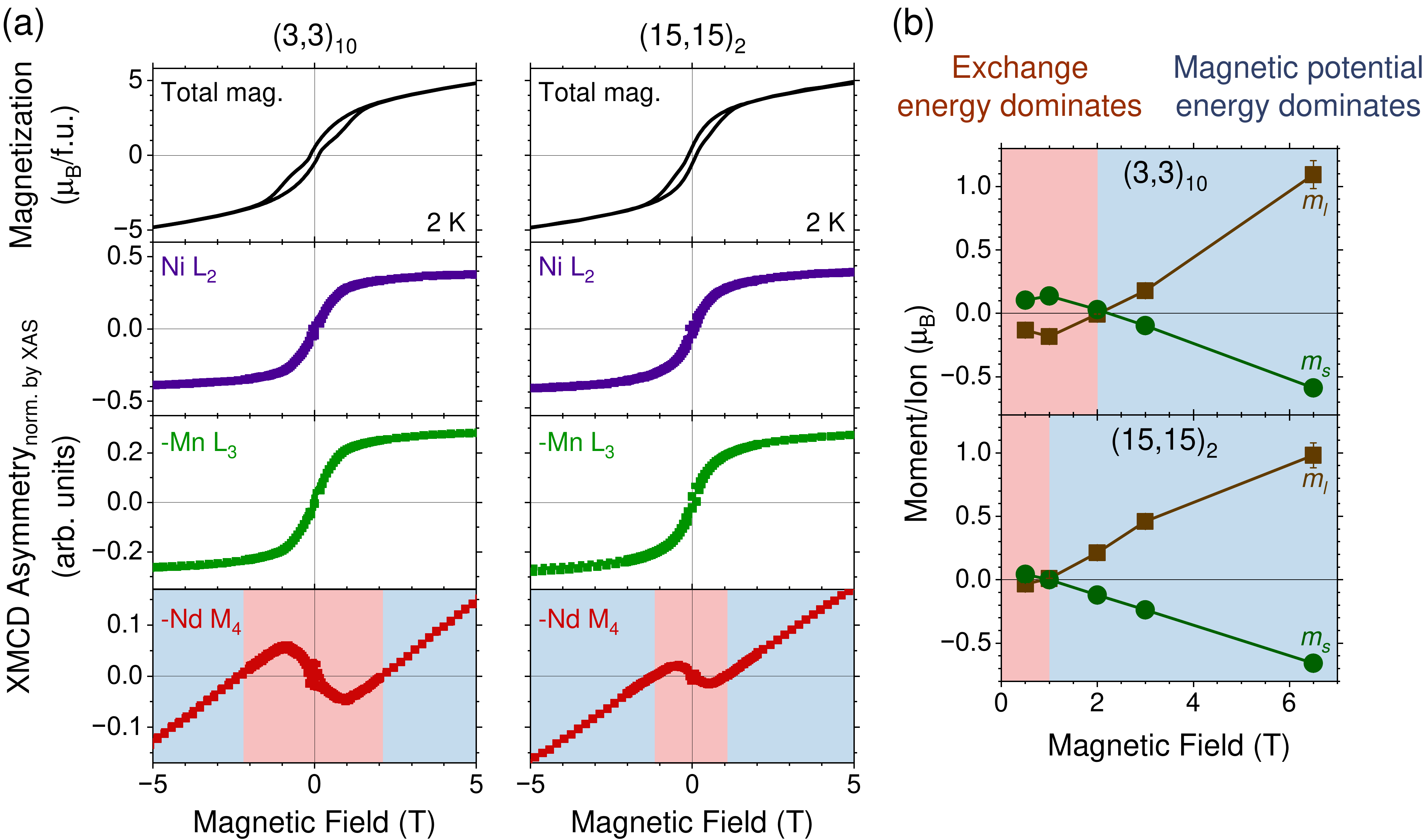}
\caption{Low-temperature magnetic characterization of a $(3,3)_{10}$ and a $(15,15)_{2}$ LNMO/NNMO superlattice. (a) $M(H)$ characterization at \SI{2}{\kelvin} for a $(3,3)_{10}$ (left) and a $(15,15)_{2}$ (right) sample. The top row shows the total magnetization measured by SQUID. The remaining rows show the XMCD asymmetry measured at the indicated absorption edges. For Ni and Mn, both samples show virtually identical ferromagnetic loops. For Nd, we find high-field paramagnetic behavior together with an inversion of the magnetic moment at low fields. The respective regions are highlighted in blue and red. The reversal is much more pronounced and occurs at higher fields in the low-periodicity $(3,3)_{10}$ sample. (b) Quantitative values for the Nd orbital and spin magnetic moments $m_l$ and $m_s$ at different applied fields extracted from XMCD spectra through the application of the sum rules for a $(3,3)_{10}$ (top) and a $(15,15)_{2}$ (bottom) superlattice. In the red low-field region, the exchange energy dominates, whereas in the blue high-field region, the magnetic potential energy dominates.}
\label{fig:3}
\end{figure*}

In the top panels of Figure \ref{fig:3}(a), we present the total magnetization vs.\ magnetic field for a low-periodicity $(3,3)_{10}$ and a high-periodicity $(15,15)_{2}$ sample measured at \SI{2}{\kelvin}. The $M(H)$ data of the respective substrates was premeasured before the superlattice growth and subtracted from the sample's $M(H)$ data. This allows us to conserve the high-field paramagnetic signal contributed by the superlattices. Both $M(H)$ loops display hysteretic openings around the origin, similar to LNMO and NNMO thin films, and high-field paramagnetic behavior, similar to NNMO thin films \cite{DeLuca2021, Spring2023}. Unlike pure NNMO films, however, the low-periodicity $(3,3)_{10}$ sample features pronounced secondary openings at $\sim\pm\SI{0.5}{\tesla}$. In the high-periodicity $(15,15)_{2}$ sample, the secondary openings are hinted at by a slight pinching of the loop. To disentangle the individual elemental magnetic contributions, we present X-ray circular magnetic dichroism (XMCD) asymmetry loops in the remaining panels of Figure \ref{fig:3}(a) recorded at the indicated absorption edges. For Ni and Mn, we find ferromagnetic behavior with saturating magnetization at high fields independent of the superlattice periodicity. For Nd, we reveal a high-field linear paramagnetic behavior, which is coupled with an intriguing reversal of the magnetic moment at low fields. The respective regions are highlighted in blue and red. This effect is strongly dependent on the superlattice periodicity and is much more pronounced on the low-periodicity $(3,3)_{10}$ sample. This goes hand in hand with the larger secondary openings in the top panel of Figure \ref{fig:3}(a). Nd XMCD asymmetry loops for additional superlattice periodicities are presented in SI Figure S5, together with data for a 30 uc NNMO film. The inversion of the Nd magnetic moment scales with the number of superlattice interfaces and finally disappears in the NNMO thin film; a ``superlattice'' with zero interfaces. This may also explain the low-temperature downturn observed in the $M(T)$ measurements in Figure \ref{fig:2}(a). Lower periodicity superlattices display a stronger downturn and a more pronounced reversal of the Nd magnetic moment. Also, the Nd reversal is strongly attenuated at temperatures above \SI{25}{\kelvin}, see SI Figure S6, coinciding with the onset of the downturn in the $M(T)$ measurements. However, this is inconsistent with the fact that a $M(T)$ downturn is also observed in the NNMO thin film, for which no Nd reversal was detected. The exact nature of the downturn, hence, remains a topic for future studies.

To quantify the behavior of the magnetic Nd sublattice, we present the Nd orbital and spin magnetic moments, $m_l$ and $m_s$, at selected positive field values in Figure \ref{fig:3}(b). The values are extracted from XMCD spectra (Nd $M_{5,4}$ edge) through application of the XMCD sum rules. The respective XMCD spectra are shown in SI Figure S7. As Hund's third rule predicts for a rare earth ion with a less than half-filled $4f$ shell, $m_l$ and $m_s$ are aligned antiparallel. The direction of the total magnetic moment is given by the dominating orbital component $m_l$. At high field values, $m_l$, and hence the total moment, are parallel to the applied field. At low fields, $m_l$ and $m_s$ are inverted, and now the total magnetic moment is antiparallel to the applied field. The inversion is strongly enhanced by the presence of superlattice interfaces. In the low-periodicity $(3,3)_{10}$ sample, which is dominated by interfacial regions, it takes place at \SI{\sim 2}{\tesla}, whereas in the high-periodicity $(15,15)_{2}$ sample, which is mostly bulk-like, it occurs at lower fields (\SI{\sim 1}{\tesla}). 

To better understand this behavior, we analyze the magnetism in LNMO/NNMO superlattices using density functional theory (DFT) calculations. We start by considering (1,1) and (3,3) superlattices and impose a ferromagnetic alignment of the Ni and Mn spins. For the Nd spins, we consider the orientations in which (1) they are aligned parallel to the Ni/Mn magnetic sublattice; (2) they are oriented parallel to each other but antiparallel to the Ni/Mn sublattice. We relax the crystal structures, imposing the lattice parameters of the substrate with these magnetic orders using DFT without spin-orbit coupling. Next, we turn on spin-orbit coupling and compute the energies of the optimized structures with corresponding magnetic orders and all the spins oriented within the $ab$-plane in which the magnetic field is applied (here, we only discuss the orientation of the spins along the $a$-axis since we found that is energetically more favorable than the $b$-axis). The energy difference $\Delta E = E_{(1)} - E_{(2)}$ between the states (1) and (2) quantifies the effective exchange field acting on the Nd moments from the Mn and Ni spins. The computed $\Delta E$ (per Nd atom) for LNMO/NNMO superlattices are presented in Table \ref{tab:deltaE}. For comparison, we also provide the corresponding value as obtained for bulk NNMO.

\begin{table}
\caption{The energy differences $\Delta E$ (in meV per Nd atom) between the states in which Nd spins are aligned parallel to each other and the Ni/Mn sublattice and the one in which they are parallel to each other, but antiparallel to the Ni/Mn sublattice computed for (1,1) and (3,3) superlattices, as well as for bulk NNMO.}
\centering
\label{tab:deltaE}
\begin{tabular}{ll}
    \toprule
    System & $\Delta E$ (meV)\\
    \midrule
    (1,1) & -273.5\\
    (3,3) & -42.8\\
    Bulk NNMO & -4.5\\
    \bottomrule
\end{tabular}
\end{table}

One can see that for both superlattices, as well as for bulk NNMO, the state in which Nd, Mn, and Ni spins are aligned ferromagnetically is lower in energy than the one in which Nd spins are antiparallel to the Ni/Mn sublattice. $m_s$ and $m_l$ of the Nd atoms are antialigned in all considered cases, as expected from the third Hund’s rule and in agreement with our XMCD measurements (see Figure \ref{fig:3}(b)). The average exchange interaction between Nd and Ni/Mn spins significantly increases as the thickness of the individual layers is reduced (superlattices with thinner layers show stronger exchange). Compared to bulk NNMO, where no interfaces are present, the exchange interaction in the (1,1) superlattice is enhanced by two orders of magnitude. This is in accordance with the absence of the Nd magnetic moment reversal in NNMO thin films, SI Figure S5. \cite{Spring2023}

\begin{figure*}
\centering
\includegraphics[width=0.6\textwidth]{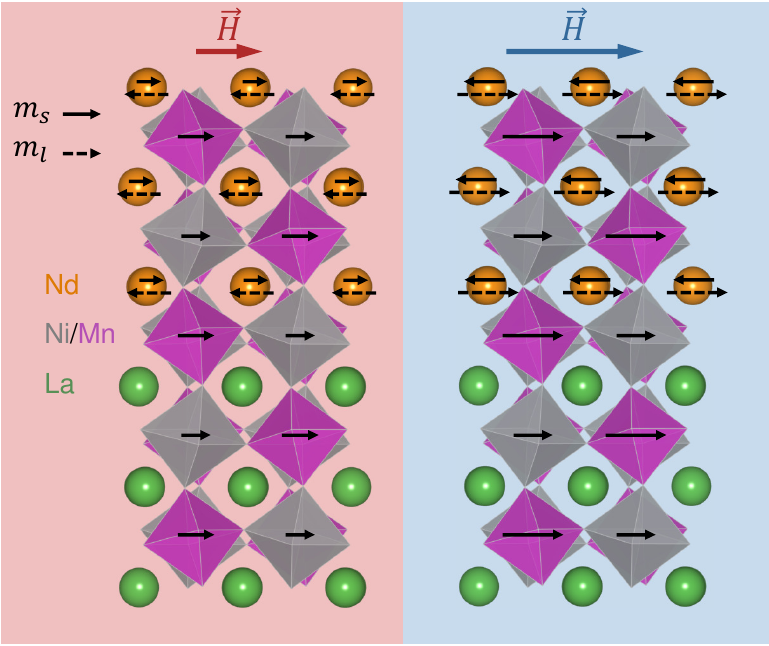}
\caption{Schematic of the magnetic moments in an LNMO/NNMO superlattice. Arrows illustrate the projection of the spin and orbital moments along the direction of the external field ($m_s$ and $m_l$, respectively). The background colors correspond to the field regions indicated in Figure \ref{fig:3}. Left: In the ground state, i.e., at zero or low applied field, the Nd spin magnetic moment aligns parallel to the Ni/Mn magnetic moment due to exchange coupling. The dominant Nd orbital magnetic moment, and hence the Nd total magnetic moment, are antiparallel. Right: At high applied fields, the driving force to align the Nd total moment in parallel to the field overcomes the Nd-Ni/Mn exchange coupling, and both the Nd orbital and spin magnetic moments flip their orientation.}
\label{fig:4}
\end{figure*}

Therefore, we present the following picture, which is illustrated in Figure \ref{fig:4}: At low fields (red region), the Nd spins $m_s$ align parallel to the Ni/Mn spin sublattice to reduce the exchange energy. Hence, the Nd orbital magnetic moments $m_l$ are antiparallel to both the Ni/Mn magnetic moments and to the external field. As the applied magnetic field increases, the Nd orbital moments $m_l$ eventually switch, so they align parallel to the field and can further grow. Simultaneously, the Nd spins $m_s$ switch as well to remain antiparallel to the orbital magnetic moments $m_l$. The magnetic field required to induce this reversal is higher for shorter-period superlattices, Figure \ref{fig:3}(b), consistent with the fact that the effective Nd-Ni/Mn exchange interactions are stronger for superlattices with thinner layers.

\section{Conclusions}
We showed how to manipulate the transition temperatures and the exchange interaction in oxide superlattices using their periodicity as a tuning knob. This was exemplified in atomically precise superlattices made of the insulating ferromagnetic double perovskites \ch{La2NiMnO6} (LNMO) and \ch{Nd2NiMnO6} (NNMO). First, we presented a magnetization vs. periodicity phase diagram, demonstrating that the respective para- to ferromagnetic transitions of LNMO and NNMO converge as the superlattice periodicity is reduced. We showed our data can be fitted to a simple Landau model that suggests a propagation of the magnetic order parameter across the superlattice interfaces. Second, we disentangled the magnetic system into a robust ferromagnetic Ni/Mn sublattice and a paramagnetic Nd sublattice with a low-field reversal of the Nd magnetic moments, showing how the Nd-Ni/Mn exchange coupling can be uniquely tuned at the interfaces.

\section{Methods}
\subsection{Growth and Structural Characterization}
The superlattices are grown by radio-frequency (RF) off-axis magnetron sputtering in a custom-built setup featuring in situ reflection high energy electron diffraction (RHEED) \cite{Podkaminer2016}. The single crystal \ch{TiO2}-terminated \ch{SrTiO3} (STO) substrates are heated to \SI{680}{\celsius} in the growth atmosphere of \SI{35}{sccm} (standard cubic centimeter per minute) Ar and \SI{10}{sccm} \ch{O2} at a pressure of \SI{0.1}{\milli\bar}. The sputter growth is carried out from stoichiometric LNMO and NNMO targets at an RF plasma power of \SI{35}{\watt}. Finally, the samples are cooled to room temperature in the growth atmosphere.

The structural quality of the superlattices is evaluated via X-ray diffraction (XRD) on a Panalytical Smartlab diffractometer using monochromatic Cu $K\alpha_1$ radiation ($\lambda = \SI{1.5406}{\angstrom}$). The surface topography is characterized by atomic force microscopy (AFM) on a Park Systems NX10.

\subsection{Magnetic Characterization}
Bulk magnetization of the samples is measured using both a Quantum Design MPMS 3 SQUID (superconducting quantum interference device) and a Quantum Design PPMS with the vibrating sample magnetometer option. For the $M(T)$ measurements, substrate contributions are removed by subtracting an STO $M(T)$ curve averaged from 15 different STO substrates. For the $M(H)$ measurements, the individual substrate is premeasured before the superlattice growth.

\subsection{Synchrotron Measurements}
To characterize the electronic and magnetic properties on an elemental level, X-ray absorption (XAS) and X-ray magnetic circular dichroism (XMCD) are performed at the X-Treme beamline at the Swiss Light Source SLS at Paul Scherrer Institute, Switzerland \cite{Piamonteze2012}. Absorption data is acquired at \SI{\sim2}{\kelvin} in grazing incidence (\SI{60}{\deg} to the sample normal) with circularly left- and right-hand polarized light. The XAS data is defined as the average of both circular polarizations. The Mn reference spectra are obtained on the following powder samples: \ch{MnCl2} (\ch{Mn^{2+}}), \ch{Mn2O3} (\ch{Mn^{3+}}) and \ch{SrMnO3} (\ch{Mn^{4+}}). For Ni we use \ch{NiO} (\ch{Ni^{2+}}) and \ch{NdNiO3} (\ch{Ni^{3+}}).

The XMCD signal is defined as the difference between the absorption signals of both polarizations. It is subsequently normalized by the most intense XAS peak. Quantitative magnetic moments are extracted from the XMCD spectra via the XMCD sum rules \cite{Carra1993}. Details on the exact procedure followed for applying the sum rules are published elsewhere \cite{Spring2023}. The XMCD asymmetry loops are obtained by measuring the XMCD signal alternatingly at an on- and off-resonance energy while sweeping the applied magnetic field.

\subsection{Electron Microscopy}
Cross-sectional electron transparent samples for STEM investigations were prepared using a FEI Helios 660 G3 UC focused ion beam (FIB) operated at acceleration voltages of 30 and \SI{5}{\kilo\volt} after deposition of C and Pt protective layers. High-angle annular dark-field scanning transmission electron microscopy (HAADF-STEM) and energy dispersive X-ray (EDX) spectroscopy were carried out using a probe-corrected FEI Titan Themis microscope equipped with ChemiSTEM technology. The microscope was operated at an accelerating voltage of \SI{300}{\kilo\volt}, probe convergence semi-angle of \SI{18}{\milli\radian}, and collecting semi-angles of \SIrange{66}{200}{\milli\radian}. The EDX spectrum image was acquired using the Velox software with a $1510\times576$ frame size, \SI{17.93}{\pico\meter} pixel size, \SI{8}{\micro\second} dwell time, and \SI{450}{\pico\ampere} probe current. The elemental maps were calculated from the EDX spectrum image using the Sr $K\alpha$, Ti $K\alpha$, La $L\alpha$, Nd $L\alpha$, Ni $K\alpha$, and Mn $K\alpha$ lines.

\subsection{Landau Model}
To theoretically model the phase transition, we used a one-dimensional Landau theory with periodic boundary conditions. Half the material is modeled as LNMO, and the other half as NNMO. Each layer of material can be treated as:

\begin{equation}
F\left(m\left(z\right),T\right)=\frac{A_{i(z)}}{4}m(z)^4+\frac{B_{i(z)}(T)}{2}m(z)^2
\end{equation}
With $B_i(T)=K(T-T^i_\text{C})$ and $i(z)=1,2$ for the two materials, $z$ the layer index, $m$ the magnetic order parameter, $T$ the temperature and $T_\text{C}^i$ the Curie temperature of each layer. The total energy for the material at temperature $T$ is:

\begin{equation}
F_T(T)=\sum_z \left[ F(m(z),T)+\frac{\xi (\nabla_z m(z))^2}{2} \right]
\end{equation}
which we can then solve using Newton's method. Since $z$ is a discrete layer number, the gradient takes the numerical form $m(z+1)-m(z)$.

The ratios $A_{i(z)}/B_{i(z)}$ are determined for each bulk material from the magnetization plot versus temperature, and the $A_1/A_2$ ratio is determined from the joint transition temperature at lower superlattice periodicities. Finally, the $\xi$ term is determined by fitting the two transition temperatures around the (5,5) periodicity.

We use $A_1=1,B_1=0.025,A_2=0.4,B_2=0.01,\xi=2$ , in arbitrary units. The MATLAB code is available under \url{https://github.com/alexandrub53/MagneticSuperlattices}.

\subsection{DFT Calculations}
All the density functional theory \cite{Hohenberg1964, Kohn1965} calculations are performed using the Vienna ab initio Simulation Package (VASP) \cite{Kresse1996} within the projector-augmented plane wave method \cite{Kresse1996, Bloechl1994}. We use generalized gradient approximation for the exchange-correlation functional in the form of Perdew, Burke, and Ernzerhof optimized for solids (PBEsol) \cite{Perdew2008}. For a more accurate treatment of localized electrons, we apply Hubbard U correction \cite{Perdew2008} of \SI{3}{\electronvolt} for $3d$ states of Mn and Ni, and \SI{9}{\electronvolt} for $4f$ states of Nd. We use a plane-wave basis set with a cutoff energy of \SI{500}{\electronvolt}. We model (1,1) and (3,3) superlattices using the 20 and 60-atom simulation cells shown in SI Figure S6(a,b) and bulk NNMO using the 20-atom cell shown in SI Figure S6(c). A $\Gamma$-centered $6\times6\times4$ Monkhorst-Pack $k$-point grid is used for reciprocal space integrals in the Brillouin zone corresponding to the 20-atom cell, while a $5\times5\times1$ $k$-grid is used for the 60-atom cell. Since all the superlattices have been grown on \ch{SrTiO3} substrate with the $c$ axis of the superlattice perpendicular to the substrate, we simulate the effects of strain by fixing the $a$ and $b$ lattice constants of the superlattices to $a\sqrt{2}$ of the substrate. The $c$ lattice constant, as well as the internal coordinates, are allowed to relax in our structural optimizations. For bulk NNMO, we optimize all the lattice constants and the ionic positions in the simulation cell. The structures are considered to be relaxed when the forces acting on the atoms are below \SI{0.01}{\electronvolt/A}. The structural relaxations are performed without spin-orbit coupling. The total energy calculations with various orientations of Nd spins are performed, including spin-orbit coupling.

\section*{Acknowledgement}
This research was supported by the Swiss National Science Foundation (SNSF) under Projects No. PP00P2\_170564 and No. 206021\_150784. J.S. and M.G. thank the Stiftung für wissenschaftliche Forschung an der UZH for their financial support under grant STWF-21-023. XAS and XMCD measurements were performed at SLS X07MA XTreme under proposal 20211683. J.S. and M.G. thank Thomas Greber for granting SQUID access and Michael Stöger-Pollach for performing electron magnetic circular dichroism measurements. Work at LIST was supported by the Luxembourg National Research Fund (FNR) through grants C20/MS/14718071/THERMODIMAT and C21/MS/15799044/FERRODYNAMICS.

J.S. and M.G. conceptualized this project and designed the experiments. The samples were grown and characterized (XRD, AFM, magnetization) by J.S. Synchrotron measurements were performed by J.S., S.J., G.D.L., and M.G., supported by C.P. A.B.G. performed the theoretical Landau calculations. N.F. and J.I.G. performed the DFT calculations. The electron microscopy was performed by M.D.R. and A.V. J.S. and M.G. wrote the first manuscript. All authors analyzed the data, discussed the results, and contributed to the final version of the paper.

\printbibliography

@article{Blasse1965,
    title = {{Ferromagnetic interactions in non-metallic perovskites}},
    year = {1965},
    journal = {Journal of Physics and Chemistry of Solids},
    author = {Blasse, G.},
    number = {12},
    month = {12},
    pages = {1969--1971},
    volume = {26},
    url = {https://linkinghub.elsevier.com/retrieve/pii/0022369765902313},
    doi = {10.1016/0022-3697(65)90231-3},
    issn = {00223697}
}

@article{DeLuca2021,
    title = {{Ferromagnetic insulating epitaxially strained La$_2$NiMnO$_6$ thin films grown by sputter deposition}},
    year = {2021},
    journal = {APL Materials},
    author = {De Luca, G. and Spring, J. and Bashir, U. and Campanini, M. and Totani, R. and Dominguez, C. and Zakharova, A. and D{\"{o}}beli, M. and Greber, T. and Rossell, M. D. and Piamonteze, C. and Gibert, M.},
    number = {8},
    month = {8},
    pages = {081111},
    volume = {9},
    publisher = {AIP Publishing LLCAIP Publishing},
    url = {https://aip.scitation.org/doi/abs/10.1063/5.0055614},
    doi = {10.1063/5.0055614},
    issn = {2166532X}
}

@article{Spring2023,
  title = {{Paramagnetic Nd sublattice and thickness-dependent ferromagnetism in ${\mathrm{Nd}}_{2}{\mathrm{NiMnO}}_{6}$ double perovskite thin films}},
  author = {Spring, Jonathan and De Luca, Gabriele and J\"ohr, Simon and Herrero-Mart\'{\i}n, Javier and Guillemard, Charles and Piamonteze, Cinthia and Ros\'ario, Carlos M. M. and Hilgenkamp, Hans and Gibert, Marta},
  journal = {Phys. Rev. Mater.},
  volume = {7},
  issue = {10},
  pages = {104407},
  numpages = {8},
  year = {2023},
  month = {Oct},
  publisher = {American Physical Society},
  doi = {10.1103/PhysRevMaterials.7.104407},
  url = {https://link.aps.org/doi/10.1103/PhysRevMaterials.7.104407}
}

@article{Dominguez2020,
    title = {{Length scales of interfacial coupling between metal and insulator phases in oxides}},
    year = {2020},
    journal = {Nature Materials},
    author = {Dom{\'{i}}nguez, Claribel and Georgescu, Alexandru B. and Mundet, Bernat and Zhang, Yajun and Fowlie, Jennifer and Mercy, Alain and Waelchli, Adrien and Catalano, Sara and Alexander, Duncan T. L. and Ghosez, Philippe and Georges, Antoine and Millis, Andrew J. and Gibert, Marta and Triscone, Jean-Marc},
    number = {11},
    month = {11},
    pages = {1182--1187},
    volume = {19},
    publisher = {Nature Research},
    url = {https://www.nature.com/articles/s41563-020-0757-x},
    doi = {10.1038/s41563-020-0757-x},
    issn = {1476-1122}
}

@article{Podkaminer2016,
    title = {{Real-time and in situ monitoring of sputter deposition with RHEED for atomic layer controlled growth}},
    year = {2016},
    journal = {APL Materials},
    author = {Podkaminer, J. P. and Patzner, J. J. and Davidson, B. A. and Eom, C. B.},
    number = {8},
    month = {8},
    pages = {086111},
    volume = {4},
    publisher = {American Institute of Physics Inc.},
    url = {http://aip.scitation.org/doi/10.1063/1.4961503},
    doi = {10.1063/1.4961503},
    issn = {2166532X}
}

@article{Piamonteze2012,
    title = {{X-Treme beamline at SLS: X-ray magnetic circular and linear dichroism at high field and low temperature}},
    year = {2012},
    journal = {Journal of Synchrotron Radiation},
    author = {Piamonteze, Cinthia and Flechsig, Uwe and Rusponi, Stefano and Dreiser, Jan and Heidler, Jakoba and Schmidt, Marcus and Wetter, Reto and Calvi, Marco and Schmidt, Thomas and Pruchova, Helena and Krempasky, Juraj and Quitmann, Christoph and Brune, Harald and Nolting, Frithjof},
    number = {5},
    month = {7},
    pages = {661--674},
    volume = {19},
    publisher = {International Union of Crystallography},
    url = {//scripts.iucr.org/cgi-bin/paper?ve5012},
    doi = {10.1107/S0909049512027847},
    issn = {0909-0495}
}

@article{Carra1993,
    title = {{X-ray circular dichroism and local magnetic fields}},
    year = {1993},
    journal = {Physical Review Letters},
    author = {Carra, Paolo and Thole, B. T. and Altarelli, Massimo and Wang, Xindong},
    number = {5},
    month = {2},
    pages = {694},
    volume = {70},
    publisher = {American Physical Society},
    url = {https://journals.aps.org/prl/abstract/10.1103/PhysRevLett.70.694},
    doi = {10.1103/PhysRevLett.70.694},
    issn = {00319007},
    pmid = {10054179}
}

@article{Dass2003,
    title = {{Oxygen stoichiometry, ferromagnetism, and transport properties of La$_{2-x}$NiMnO$_{6-\delta}$}},
    year = {2003},
    journal = {Physical Review B},
    author = {Dass, I. and Yan, J. Q. and Goodenough, B.},
    number = {6},
    month = {8},
    pages = {064415},
    volume = {68},
    publisher = {American Physical Society},
    url = {https://journals.aps.org/prb/abstract/10.1103/PhysRevB.68.064415},
    doi = {10.1103/PhysRevB.68.064415},
    issn = {1550235X}
}

@article{Booth2009,
    title = {{An investigation of structural, magnetic and dielectric properties of R$_2$NiMnO$_6$ (R=rare earth, Y)}},
    year = {2009},
    journal = {Materials Research Bulletin},
    author = {Booth, R.J. and Fillman, R. and Whitaker, H. and Nag, Abanti and Tiwari, R.M. and Ramanujachary, K.V. and Gopalakrishnan, J. and Lofland, S.E.},
    number = {7},
    month = {7},
    pages = {1559--1564},
    volume = {44},
    url = {https://linkinghub.elsevier.com/retrieve/pii/S0025540809000609},
    doi = {10.1016/j.materresbull.2009.02.003},
    issn = {00255408}
}

@article{Hwang2012,
    title = {{Emergent phenomena at oxide interfaces}},
    year = {2012},
    journal = {Nature Materials},
    author = {Hwang, H. Y. and Iwasa, Y. and Kawasaki, M. and Keimer, B. and Nagaosa, N. and Tokura, Y.},
    number = {2},
    month = {2},
    pages = {103--113},
    volume = {11},
    publisher = {Nature Publishing Group},
    url = {https://www.nature.com/articles/nmat3223},
    doi = {10.1038/nmat3223},
    issn = {1476-1122}
}

@article{kanamori1959,
    title = {{Superexchange interaction and symmetry properties of electron orbitals}},
    year = {1959},
    journal = {Journal of Physics and Chemistry of Solids},
    author = {Kanamori, Junjiro},
    number = {2},
    pages = {87--98},
    volume = {10},
    url = {https://www.sciencedirect.com/science/article/pii/0022369759900617},
    doi = {https://doi.org/10.1016/0022-3697(59)90061-7},
    issn = {0022-3697}
}

@article{Zubko2011,
    title = {{Interface Physics in Complex Oxide Heterostructures}},
    year = {2011},
    journal = {Annual Review of Condensed Matter Physics},
    author = {Zubko, Pavlo and Gariglio, Stefano and Gabay, Marc and Ghosez, Philippe and Triscone, Jean-Marc},
    number = {1},
    month = {3},
    pages = {141--165},
    volume = {2},
    publisher = {Annual Reviews},
    url = {http://www.annualreviews.org/doi/10.1146/annurev-conmatphys-062910-140445},
    doi = {10.1146/annurev-conmatphys-062910-140445},
    issn = {1947-5454},
}

@article{Catalan2000,
    title = {{Metal-insulator transitions in NdNiO$_3$ thin films}},
    year = {2000},
    journal = {Physical Review B},
    author = {Catalan, G. and Bowman, R. M. and Gregg, J. M.},
    number = {12},
    month = {9},
    pages = {7892--7900},
    volume = {62},
    publisher = {American Physical Society},
    url = {https://link.aps.org/doi/10.1103/PhysRevB.62.7892},
    doi = {10.1103/PhysRevB.62.7892},
    issn = {0163-1829}
}

@article{Tiwari2002,
    title = {{Strain-induced tuning of metal-insulator transition in NdNiO$_3$}},
    year = {2002},
    journal = {Applied Physics Letters},
    author = {Tiwari, Ashutosh and Jin, C. and Narayan, J.},
    number = {21},
    month = {5},
    pages = {4039--4041},
    volume = {80},
    publisher = {AIP Publishing},
    url = {/aip/apl/article/80/21/4039/512060/Strain-induced-tuning-of-metal-insulator},
    doi = {10.1063/1.1480475},
    issn = {0003-6951}
}

@article{Liu2010,
    title = {{Strain-mediated metal-insulator transition in epitaxial ultrathin films of NdNiO$_3$}},
    year = {2010},
    journal = {Applied Physics Letters},
    author = {Liu, Jian and Kareev, M. and Gray, B. and Kim, J. W. and Ryan, P. and Dabrowski, B. and Freeland, J. W. and Chakhalian, J.},
    number = {23},
    volume = {96},
    pages = {233110},
    doi = {10.1063/1.3451462},
    issn = {00036951}
}

@article{Catalano2014,
    title = {{Electronic transitions in strained SmNiO$_3$ thin films}},
    year = {2014},
    journal = {APL Materials},
    author = {Catalano, S. and Gibert, M. and Bisogni, V. and Peil, O. E. and He, F. and Sutarto, R. and Viret, M. and Zubko, P. and Scherwitzl, R. and Georges, A. and Sawatzky, G. A. and Schmitt, T. and Triscone, J. M.},
    number = {11},
    month = {11},
    pages = {116110},
    volume = {2},
    publisher = {American Institute of Physics Inc.},
    doi = {10.1063/1.4902138},
    issn = {2166532X}
}

@article{Ohtomo2004,
    title = {{A high-mobility electron gas at the LaAlO$_3$/SrTiO$_3$ heterointerface}},
    year = {2004},
    journal = {Nature},
    author = {Ohtomo, A. and Hwang, H. Y.},
    pages = {423--426},
    volume = {427},
    doi = {10.1038/nature02308},
    issn = {00280836}
}

@article{Reyren2007,
    title = {{Superconducting interfaces between insulating oxides}},
    year = {2007},
    journal = {Science},
    author = {Reyren, N. and Thiel, S. and Caviglia, A. D. and Fitting Kourkoutis, L. and Hammerl, G. and Richter, C. and Schneider, C. W. and Kopp, T. and R{\"{u}}etschi, A. S. and Jaccard, D. and Gabay, M. and Muller, D. A. and Triscone, J. M. and Mannhart, J.},
    number = {5842},
    month = {8},
    pages = {1196--1199},
    volume = {317},
    doi = {10.1126/science.1146006},
    issn = {00368075}
}

@article{Dominguez2023,
  title = {Coupling of magnetic phases at nickelate interfaces},
  author = {Dom\'{\i}nguez, C. and Fowlie, J. and Georgescu, A. B. and Mundet, B. and Jaouen, N. and Viret, M. and Suter, A. and Millis, A. J. and Salman, Z. and Prokscha, T. and Gibert, M. and Triscone, J.-M.},
  journal = {Phys. Rev. Mater.},
  volume = {7},
  issue = {6},
  pages = {065002},
  numpages = {10},
  year = {2023},
  month = {Jun},
  publisher = {American Physical Society},
  doi = {10.1103/PhysRevMaterials.7.065002},
  url = {https://link.aps.org/doi/10.1103/PhysRevMaterials.7.065002}
}

@article{Mundy2016,
    title = {{Atomically engineered ferroic layers yield a room-temperature magnetoelectric multiferroic}},
    year = {2016},
    journal = {Nature},
    author = {Mundy, Julia A. and Brooks, Charles M. and Holtz, Megan E. and Moyer, Jarrett A. and Das, Hena and R{\'{e}}bola, Alejandro F. and Heron, John T. and Clarkson, James D. and Disseler, Steven M. and Liu, Zhiqi and Farhan, Alan and Held, Rainer and Hovden, Robert and Padgett, Elliot and Mao, Qingyun and Paik, Hanjong and Misra, Rajiv and Kourkoutis, Lena F. and Arenholz, Elke and Scholl, Andreas and Borchers, Julie A. and Ratcliff, William D. and Ramesh, Ramamoorthy and Fennie, Craig J. and Schiffer, Peter and Muller, David A. and Schlom, Darrell G.},
    number = {7621},
    month = {9},
    pages = {523--527},
    volume = {537},
    publisher = {Nature Publishing Group},
    doi = {10.1038/nature19343},
    issn = {14764687}
}

@article{Hohenberg1964,
  title = {Inhomogeneous Electron Gas},
  author = {Hohenberg, P. and Kohn, W.},
  journal = {Phys. Rev.},
  volume = {136},
  issue = {3B},
  pages = {B864--B871},
  numpages = {0},
  year = {1964},
  month = {Nov},
  publisher = {American Physical Society},
  doi = {10.1103/PhysRev.136.B864},
  url = {https://link.aps.org/doi/10.1103/PhysRev.136.B864}
}

@article{Kohn1965,
  title = {Self-Consistent Equations Including Exchange and Correlation Effects},
  author = {Kohn, W. and Sham, L. J.},
  journal = {Phys. Rev.},
  volume = {140},
  issue = {4A},
  pages = {A1133--A1138},
  numpages = {0},
  year = {1965},
  month = {Nov},
  publisher = {American Physical Society},
  doi = {10.1103/PhysRev.140.A1133},
  url = {https://link.aps.org/doi/10.1103/PhysRev.140.A1133}
}

@article{Kresse1996,
  title = {Efficient iterative schemes for ab initio total-energy calculations using a plane-wave basis set},
  author = {Kresse, G. and Furthm\"uller, J.},
  journal = {Phys. Rev. B},
  volume = {54},
  issue = {16},
  pages = {11169--11186},
  numpages = {0},
  year = {1996},
  month = {Oct},
  publisher = {American Physical Society},
  doi = {10.1103/PhysRevB.54.11169},
  url = {https://link.aps.org/doi/10.1103/PhysRevB.54.11169}
}

@article{Bloechl1994,
  title = {Projector augmented-wave method},
  author = {Bl\"ochl, P. E.},
  journal = {Phys. Rev. B},
  volume = {50},
  issue = {24},
  pages = {17953--17979},
  numpages = {0},
  year = {1994},
  month = {Dec},
  publisher = {American Physical Society},
  doi = {10.1103/PhysRevB.50.17953},
  url = {https://link.aps.org/doi/10.1103/PhysRevB.50.17953}
}

@article{Perdew2008,
  title = {Restoring the Density-Gradient Expansion for Exchange in Solids and Surfaces},
  author = {Perdew, John P. and Ruzsinszky, Adrienn and Csonka, G\'abor I. and Vydrov, Oleg A. and Scuseria, Gustavo E. and Constantin, Lucian A. and Zhou, Xiaolan and Burke, Kieron},
  journal = {Phys. Rev. Lett.},
  volume = {100},
  issue = {13},
  pages = {136406},
  numpages = {4},
  year = {2008},
  month = {Apr},
  publisher = {American Physical Society},
  doi = {10.1103/PhysRevLett.100.136406},
  url = {https://link.aps.org/doi/10.1103/PhysRevLett.100.136406}
}

@article{Ramesh2019,
  title={Creating emergent phenomena in oxide superlattices},
  author={Ramesh, Ramamoorthy and Schlom, Darrell G},
  journal={Nature Reviews Materials},
  volume={4},
  number={4},
  pages={257--268},
  year={2019},
  publisher={Nature Publishing Group UK London}
}

@article{Bhattacharya2014,
  title={Magnetic oxide heterostructures},
  author={Bhattacharya, Anand and May, Steven J},
  journal={Annual Review of Materials Research},
  volume={44},
  pages={65--90},
  year={2014},
  publisher={Annual Reviews}
}

@article{Takahashi2001,
  title={Interface ferromagnetism in oxide superlattices of CaMnO$_3$/CaRuO$_3$},
  author={Takahashi, KS and Kawasaki, M and Tokura, Y},
  journal={Applied Physics Letters},
  volume={79},
  number={9},
  pages={1324--1326},
  year={2001},
  publisher={American Institute of Physics}
}

@article{Novoselov2016,
author = {K. S. Novoselov  and A. Mishchenko  and A. Carvalho  and A. H. Castro Neto },
title = {2D materials and van der Waals heterostructures},
journal = {Science},
volume = {353},
number = {6298},
pages = {aac9439},
year = {2016},
doi = {10.1126/science.aac9439},
URL = {https://www.science.org/doi/abs/10.1126/science.aac9439},
eprint = {https://www.science.org/doi/pdf/10.1126/science.aac9439}
}

@article{Geim2013,
  title={Van der Waals heterostructures},
  author={Geim, Andre K and Grigorieva, Irina V},
  journal={Nature},
  volume={499},
  number={7459},
  pages={419--425},
  year={2013},
  publisher={Nature Publishing Group UK London}
}

@article{Alferov1998,
  title={The history and future of semiconductor heterostructures},
  author={Alferov, Zh I},
  journal={Semiconductors},
  volume={32},
  pages={1--14},
  year={1998},
  publisher={Springer}
}

@article{Liao2016,
  title={Controlled lateral anisotropy in correlated manganite heterostructures by interface-engineered oxygen octahedral coupling},
  author={Liao, Zhaoliang and Huijben, Mark and Zhong, Z and Gauquelin, N and Macke, S and Green, RJ and Van Aert, S and Verbeeck, J and Van Tendeloo, G and Held, K and Sawatzky, G A and Koster, G and Rijnders, G},
  journal={Nature materials},
  volume={15},
  number={4},
  pages={425--431},
  year={2016},
  publisher={Nature Publishing Group UK London}
}

@article{Flint2017,
  title={Role of polar compensation in interfacial ferromagnetism of LaNiO$_3$/CaMnO$_3$ superlattices},
  author={Flint, Charles L and Jang, Hoyoung and Lee, J-S and N'Diaye, Alpha T and Shafer, Padraic and Arenholz, Elke and Suzuki, Yuri},
  journal={Physical Review Materials},
  volume={1},
  number={2},
  pages={024404},
  year={2017},
  publisher={APS}
}

@article{Grutter2016,
  title={Interfacial symmetry control of emergent ferromagnetism at the nanoscale},
  author={Grutter, Alexander J and Vailionis, Arturas and Borchers, Julie A and Kirby, Brian J and Flint, CL and He, Chunyong and Arenholz, Elke and Suzuki, Yuri},
  journal={Nano Letters},
  volume={16},
  number={9},
  pages={5647--5651},
  year={2016},
  publisher={ACS Publications}
}

@article{DeLuca2022,
    title = {{Top-Layer Engineering Reshapes Charge Transfer at Polar Oxide Interfaces}},
    year = {2022},
    journal = {Advanced Materials},
    author = {De Luca, Gabriele and Spring, Jonathan and Kaviani, Moloud and J{\"{o}}hr, Simon and Campanini, Marco and Zakharova, Anna and Guillemard, Charles and Herrero-Martin, Javier and Erni, Rolf and Piamonteze, Cinthia and Rossell, Marta D and Aschauer, Ulrich and Gibert, Marta},
    number = {36},
    pages = {2203071},
    volume = {34},
    publisher = {John Wiley {\&} Sons, Ltd},
    url = {https://onlinelibrary.wiley.com/doi/full/10.1002/adma.202203071 https://onlinelibrary.wiley.com/doi/abs/10.1002/adma.202203071 https://onlinelibrary.wiley.com/doi/10.1002/adma.202203071},
    doi = {10.1002/ADMA.202203071},
    issn = {1521-4095}
}

@article{Golalikhani2018,
  title={Nature of the metal-insulator transition in few-unit-cell-thick LaNiO$_3$ films},
  author={Golalikhani, M and Lei, Q and Chandrasena, R U and Kasaei, L and Park, H and Bai, J and Orgiani, P and Ciston, J and Sterbinsky, G E and Arena, D A and Shafer, P and Arenholz, E and Davidson, B A and Millis, A J and Gray, A X and XI, X X},
  journal={Nature communications},
  volume={9},
  number={1},
  pages={2206},
  year={2018},
  publisher={Nature Publishing Group UK London}
}

@article{Kan2016,
  title={Tuning magnetic anisotropy by interfacially engineering the oxygen coordination environment in a transition metal oxide},
  author={Kan, Daisuke and Aso, Ryotaro and Sato, Riko and Haruta, Mitsutaka and Kurata, Hiroki and Shimakawa, Yuichi},
  journal={Nature materials},
  volume={15},
  number={4},
  pages={432--437},
  year={2016},
  publisher={Nature Publishing Group UK London}
}

@article{Kumah2014,
  title={Tuning the structure of nickelates to achieve two-dimensional electron conduction},
  author={Kumah, Divine P and Disa, Ankit S and Ngai, Joseph H and Chen, Hanghui and Malashevich, Andrei and Reiner, James W and Ismail-Beigi, Sohrab and Walker, Frederick J and Ahn, Charles H},
  journal={Advanced Materials},
  volume={26},
  number={12},
  pages={1935--1940},
  year={2014}
}

@article{Huang2018,
  title={Interface Engineering and Emergent Phenomena in Oxide Heterostructures},
  author={Huang, Zhen and Ariando and Wang, X Renshaw and Rusydi, Andrivo and Chen, Jingsheng and Yang, Hyunsoo and Venkatesan, Thirumalai},
  journal={Advanced Materials},
  volume={30},
  number={47},
  pages={1802439},
  year={2018},
  publisher={Wiley Online Library}
}

\newpage
\section*{Supporting Information}
\renewcommand{\thefigure}{S\arabic{figure}}
\setcounter{figure}{0}
\renewcommand{\thetable}{S\arabic{table}}
\setcounter{table}{0}

\begin{figure}[H]
	\centering
	\includegraphics[width=0.8\textwidth]{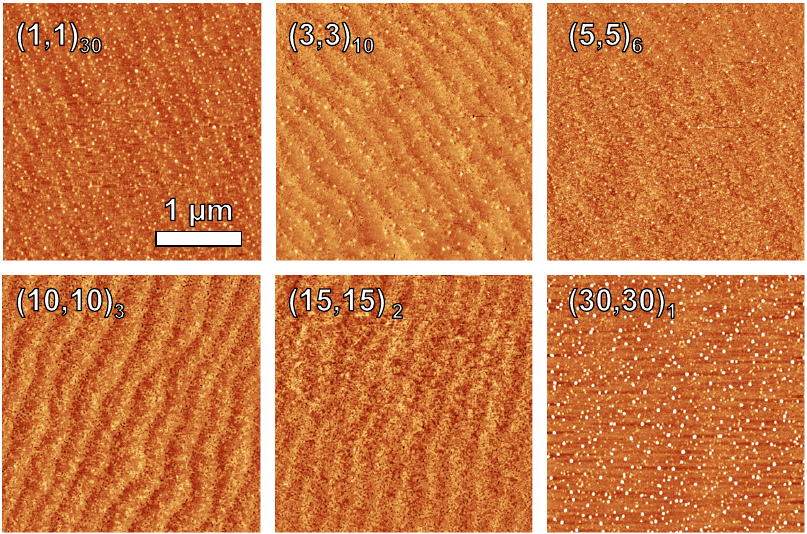}
	\caption{AFM topography images for selected superlattices. All samples except the $(30,30)_1$ show the step-and-terrace morphology inherited from the substrate.}
	\label{SIfig:1}
\end{figure}

\begin{figure}[H]
	\centering
	\includegraphics[width=\textwidth]{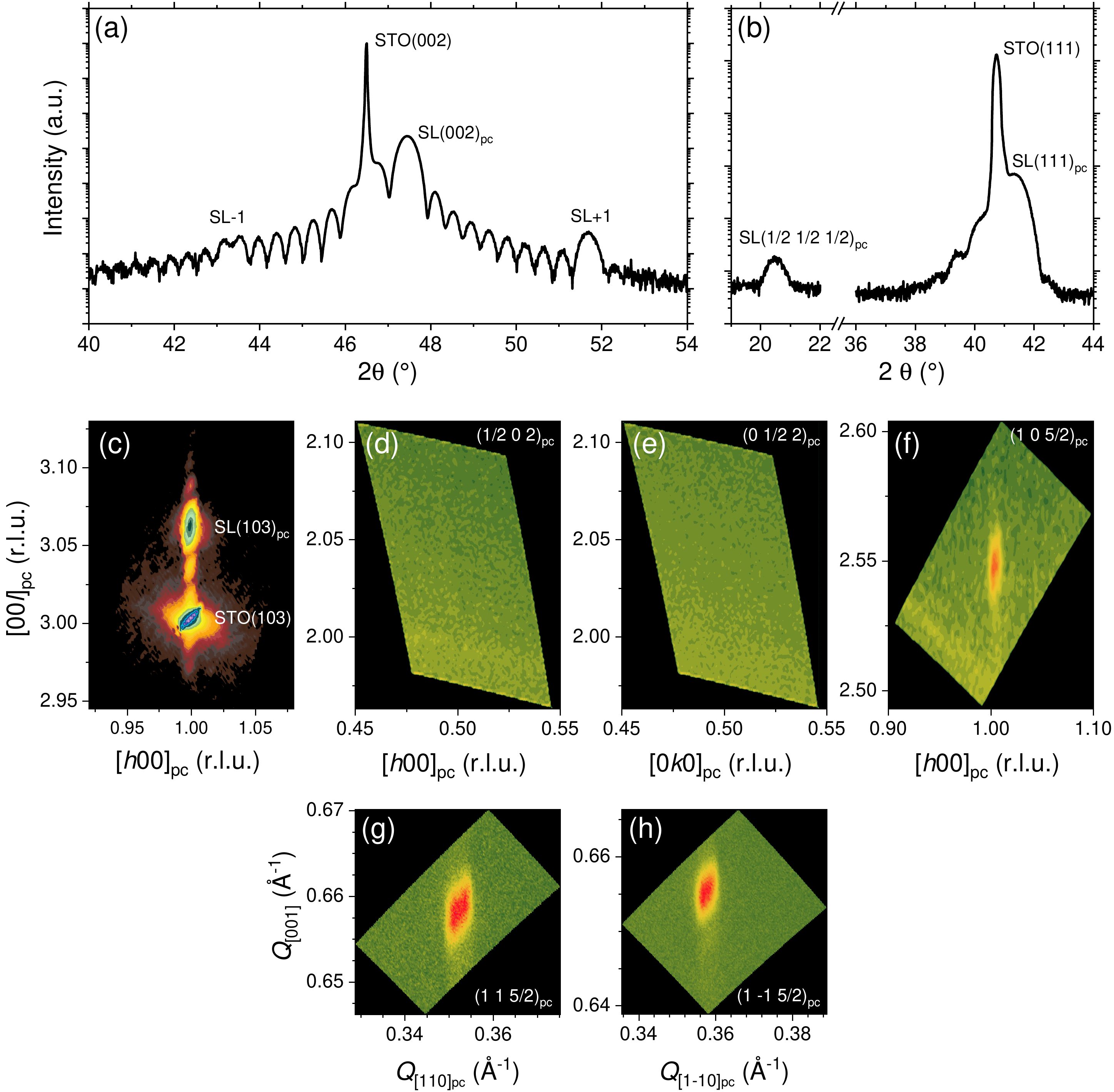}
	\caption{XRD on a $(3,3)_{10}$ superlattice. \textbf{(a)} Around the STO (002) substrate peak. \textbf{(b)} Around the STO (111) substrate peak. The emergence of a (\nicefrac{1}{2} \nicefrac{1}{2} \nicefrac{1}{2})\textsubscript{pc} diffraction peak alludes to the Ni/Mn rock salt ordering on the B-site. \textbf{(c)} RSM around the STO (103) substrate peak. The superlattice (103)\textsubscript{pc} peak is found at the same in-plane reciprocal lattice value $h$, indicating coherent epitaxial strain. \textbf{(d-f)} RSMs around the superlattice's (\nicefrac{1}{2} 0 2)\textsubscript{pc}, (0 \nicefrac{1}{2} 2)\textsubscript{pc}, and (1 0 \nicefrac{5}{2})\textsubscript{pc} diffraction conditions. A doubling of the unit cell in the (001)\textsubscript{pc} direction (f) indicates the out-of-plane orientation of the orthorhombic c-axis. \textbf{(g,h)} RSM around (1 1 \nicefrac{5}{2})\textsubscript{pc} and (1 -1 \nicefrac{5}{2})\textsubscript{pc} reveals two in-plane orientations for the superlattice.}
	\label{SIfig:2}
\end{figure}

\begin{figure}[H]
	\centering
	\includegraphics[width=\textwidth]{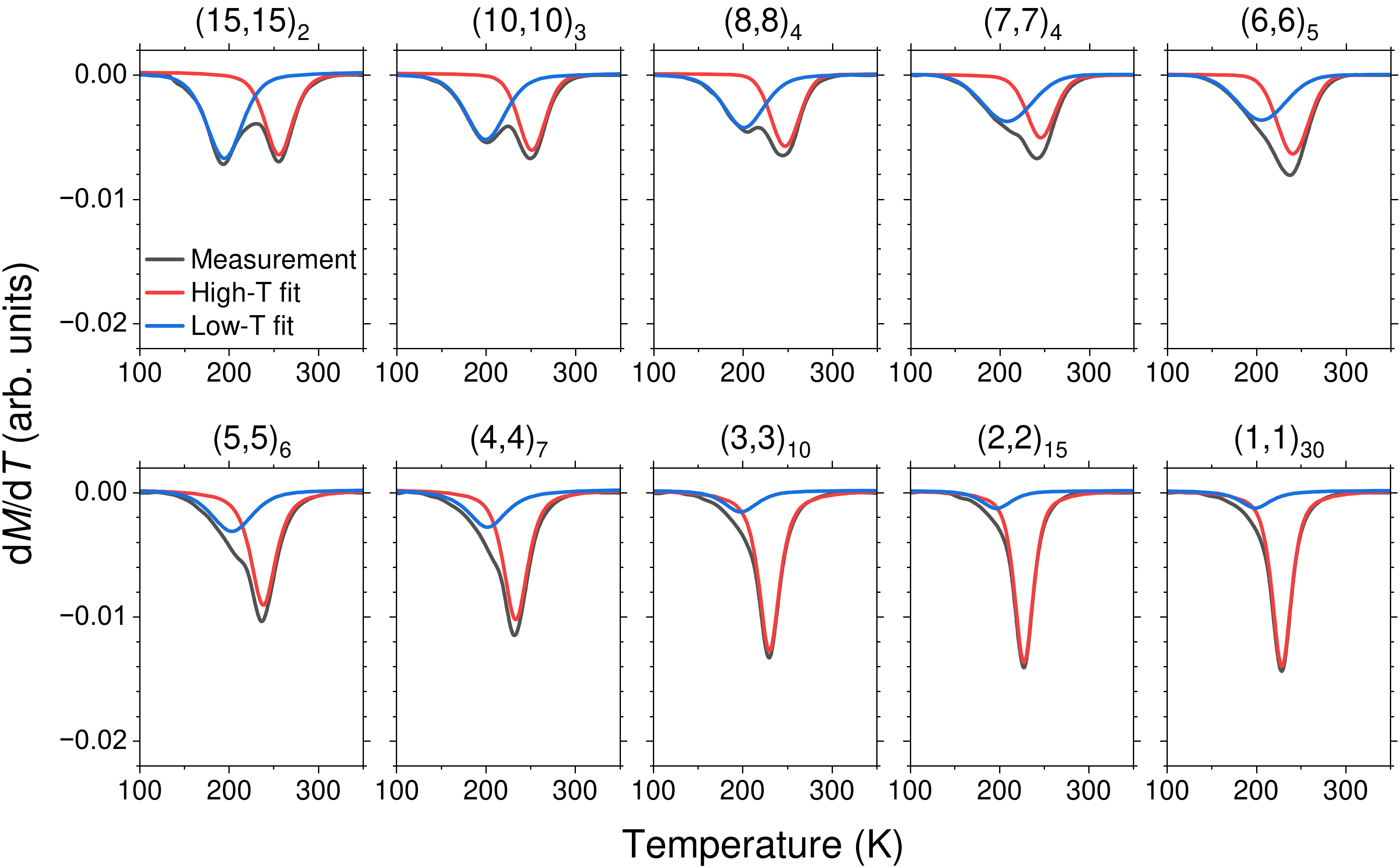}
	\caption{$\nicefrac{\text{d}M}{\text{d}T}$ data and Pseudo-Voigt fits for superlattices of selected periodicities.}
	\label{SIfig:3}
\end{figure}

\begin{figure}[H]
	\centering
	\includegraphics[width=0.5\textwidth]{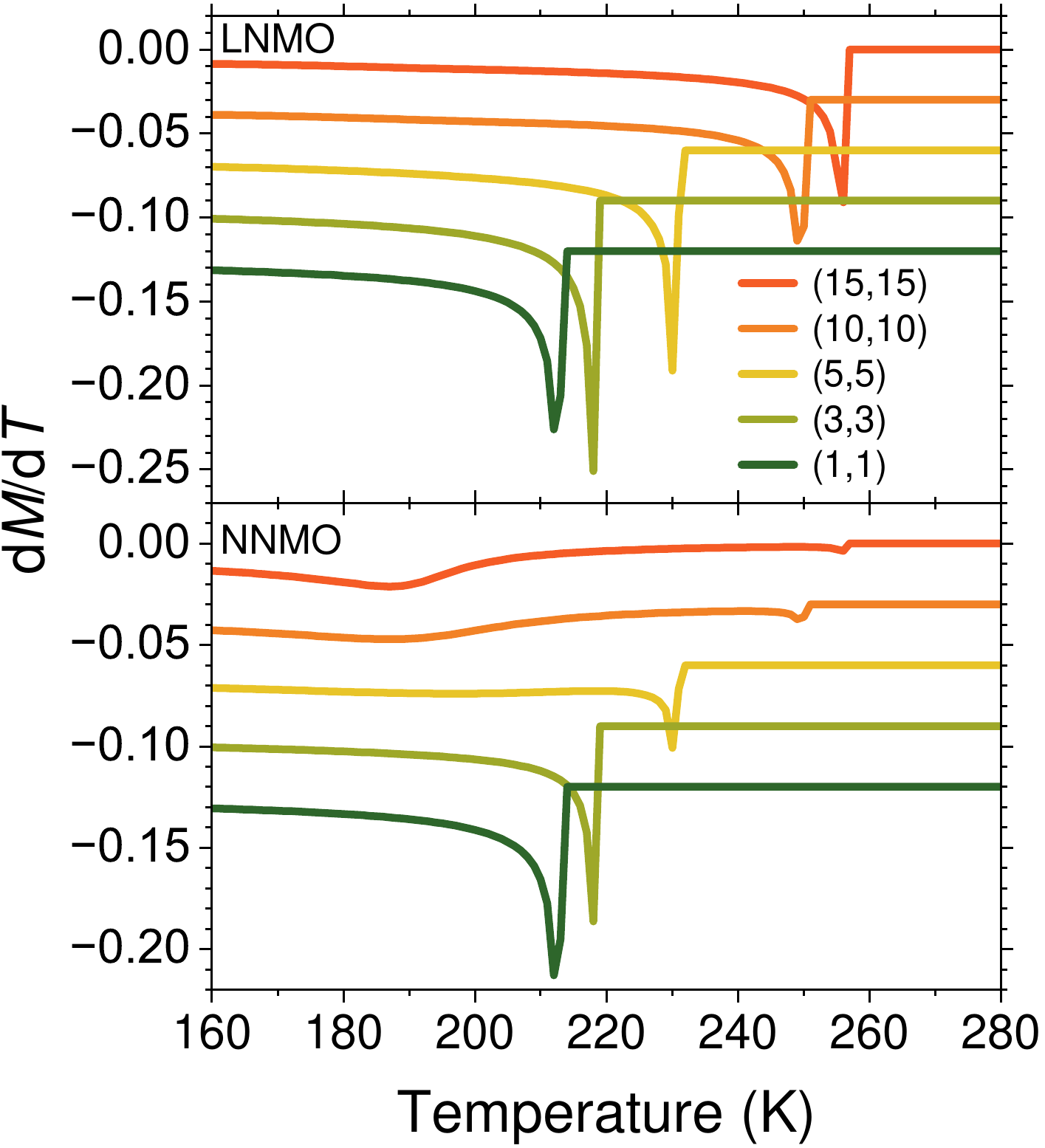}
	\caption{Calculated $\nicefrac{\text{d}M}{\text{d}T}$ data for superlattices of selected periodicities separated for LNMO and NNMO. The presented data corresponds to the center position of a LNMO or NNMO slab in a given superlattice.}
	\label{SIfig:4}
\end{figure}

\begin{figure}[H]
	\centering
	\includegraphics[width=\textwidth]{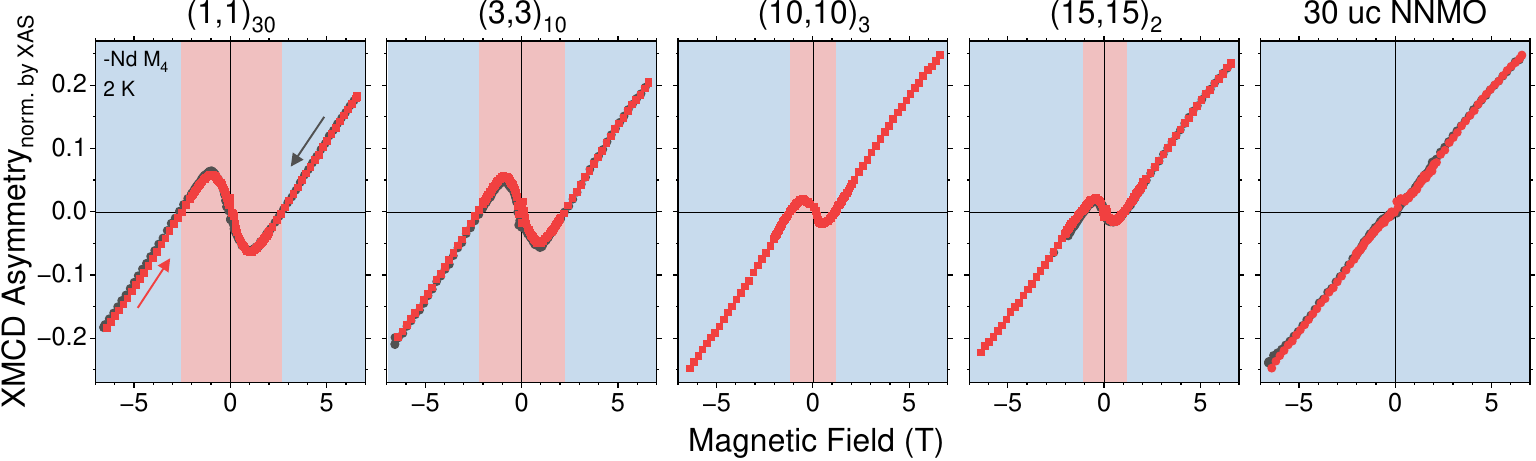}
	\caption{XMCD asymmetry loops recorded at the Nd $M_4$ absorption edge at \SI{2}{\kelvin} for LNMO/NNMO superlattices of different periodicities and a 30 uc NNMO thin film. The background colors correspond to the high- and low-field regions, as introduced in Figure 3 of the main text.}
	\label{SIfig:5}
\end{figure}

\begin{figure}[H]
	\centering
	\includegraphics[width=0.7\textwidth]{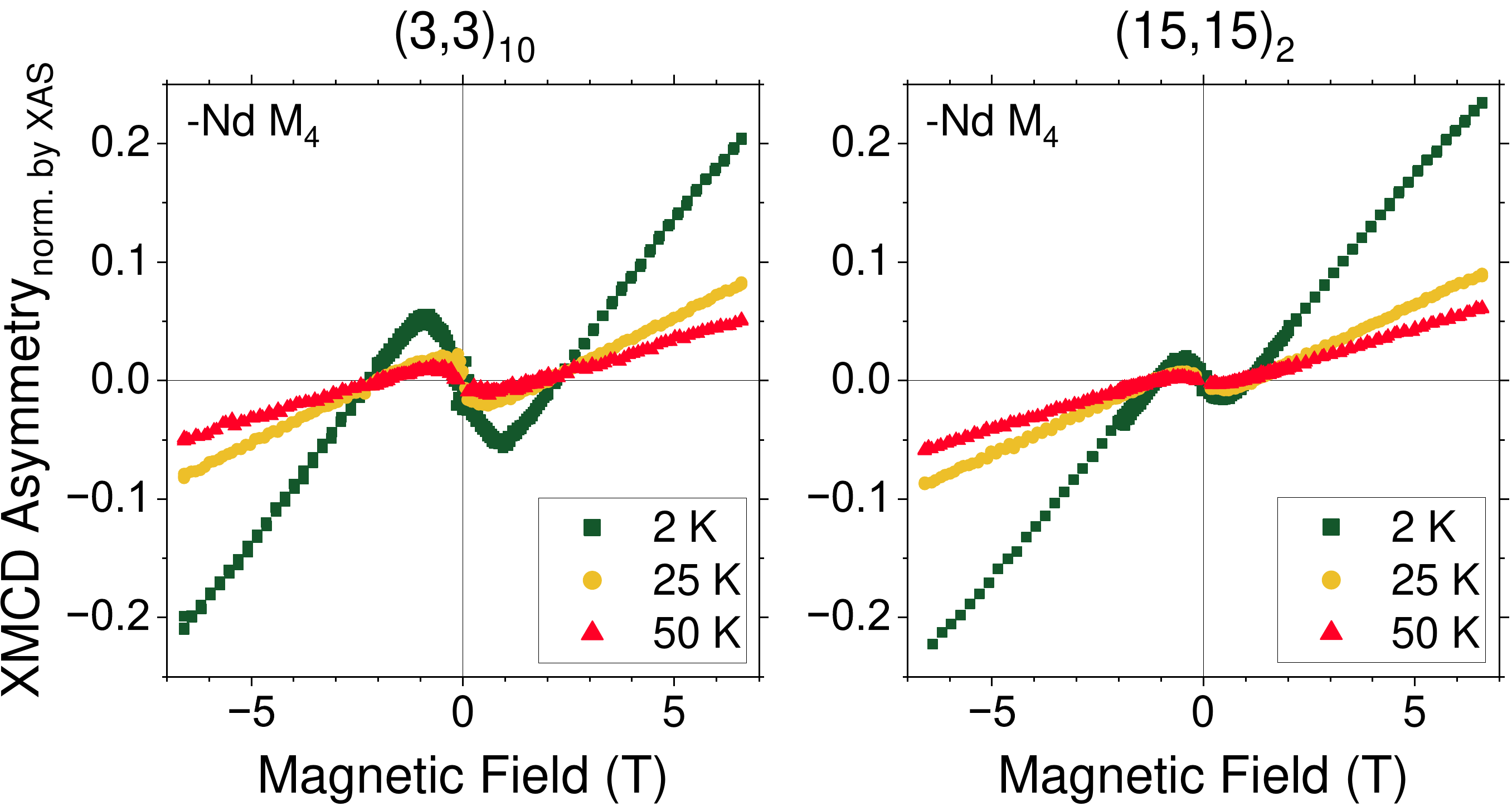}
	\caption{XMCD asymmetry loops recorded at the Nd $M_4$ absorption edge for $(3,3)_{10}$ and a $(15,15)_2$ superlattice recorded at different temperatures.}
	\label{SIfig:6}
\end{figure}

\begin{figure}[H]
	\centering
	\includegraphics[width=\textwidth]{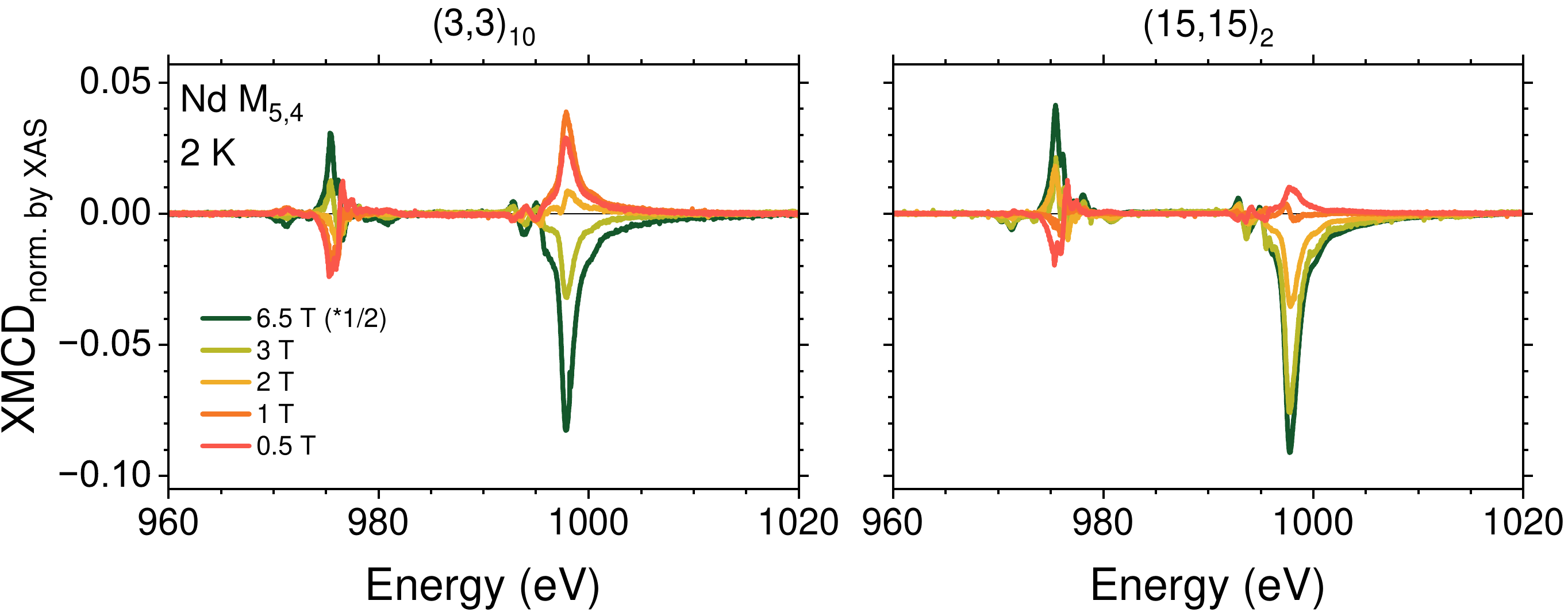}
	\caption{XMCD analysis at the Nd M$_{5,4}$ absorption edge at \SI{2}{\kelvin} and different applied magnetic fields for a $(3,3)_{10}$ and a $(15,15)_2$ superlattice.}
	\label{SIfig:7}
\end{figure}

\end{document}